\documentstyle[12pt]{article}
\newcounter{mycount}

\newcommand{\be}{\begin{eqnarray}}
\newcommand{\ee}{\end{eqnarray}}

\newcommand\ie {{\it i.e. }}
\newcommand\eg {{\it e.g. }}

\newcommand\half{\frac 1 2 }

\newcommand\noi{\noindent}

\begin{document}
\bibliographystyle{[hansson.paper]nphys}

\newcommand\go{\omega}

\centerline{\Large\bf The  Calogero Model -  Anyonic Representation, }
\vskip 2mm
\centerline{\Large\bf Fermionic Extension and Supersymmetry}
\vspace* {-40 mm}
\begin{flushright} USITP-92-14 \\
G{\"o}teborg ITP-92-53 \\ January 1993
\end{flushright}

\vskip 0.9in
\centerline{L. Brink$^\circ$, T.H.\ Hansson$^\dagger$, S. Konstein$^\star$ and
M. A. Vasiliev$^\star$ }

\vskip 2cm
\centerline{\bf ABSTRACT}
\vskip 3mm
We discuss several applications and extensions of our previous
operator solution of the $N$-body Calogero problem, \ie N particles in
1 dimension subject to a two-body interaction of the form $\half
\sum_{i,j}[ (x_i - x_j)^2 + g/ {(x_i - x_j)^2}]$.  Using a complex
representation of the deformed Heisenberg algebra underlying the
Calogero model, we explicitly establish the equivalence between this
system and anyons in the lowest Landau level. A construction based on
supersymmetry is used to extend our operator method to include
fermions, and we obtain an explicit solution of the supersymmetric
Calogero model constructed by Freedman and Mende. We also show how the
dynamical $OSp(2;2)$ supersymmetry is realized by bilinears of
modified creation and annihilation operators, and how to construct a
supersymmetic extension of the deformed Heisenberg algebra.

\vfill\noi
$^\circ$Institute of Theoretical Physics, S-41296 G{\"o}teborg,
Sweden.
\vskip 2mm\noi
$^\dagger$Institute of Theoretical Physics, University of Stockholm,
Vanadisv\"agen 9, \\ S-113 46 Stockholm, Sweden.
\vskip 2mm\noi
$^\star$Department of Theoretical Physics, P. N. Lebedev Physical
Institute,
\\
117924 Leninsky Prospect 53, Moscow, Russia.
\vskip 3mm \noi
$^{\circ\, \dagger}$Supported by the Swedish Natural Science Research
Council
\eject

\renewcommand{\theequation}{1.\arabic{equation}}
\setcounter{equation}{0}
\noi {\large\bf 1. Introduction and Summary}
\vskip 3mm

The Calogero model is a quantum mechanical system of $N$ particles on
a line interacting via the two-body potential $\go^2 x^2 +\nu (\nu
-1)x^{-2}$, where $\omega$ is the harmonic oscillator frequency and
$\nu(\nu -1)$ the coupling constant.  As pointed out by
Calogero\cite{calo1,calo2} in his 1971 paper, this model has some very
remarkable properties.  First, in order to be normalizable, all wave
fucntions must, up to a common factor, be either totally symmetric or
totally antisymmetric. Secondly, the energy spectrum is that of $N$
bosons or fermions interacting via harmonic forces only, but with a
total energy shift proportional to $\nu$. Although the spectrum is
known, it has been proved difficult to construct the corresponding
wave functions.  With rather laborious techniques, the wave functions
up to the five-particles one were constructed\cite{calo2,pere3,gamb1},
but in a recent paper\cite{brin3} we found all $N$ particle
eigenwavefunctions using an operator formulation.

The Calogero model is a prime example of a solvable low-dimensional
model, with interesting physical applications as well as deep
connections to various branches of mathematics.  In fact, it has been
shown to be but one in a large family of two dimensional ($2d$)
quantum integrable models (for a review see
\eg \cite{olsh2}). In
particular, it is expected to have close links to $2d$ conformal
models.  This is indicated by the observations that the model is
closely related to the matrix models \cite{kaza1,olsh3} and that the
differential operators which are central in our treatment of the model
(see also \cite{poly6}) appear in the decoupling equations in certain
formulations of conformal models \cite{gome1,gors1,make1,semi1}.
Another intriguing connection is to the higher-spin gauge theories in
3 and 4 space-time dimensions (see
\cite{frad3,blen1,frad4,vasi2,vasi3} and references therein) based on
a certain class of infinite-dimensional symmetry algebras, higher-spin
algebras, introduced in
\cite{Frad5,vasi4,kons1,vasi5,vasi1}.
In \cite{brin3} it was observed that a version of higher-spin algebras
investigated in
\cite{vasi5,vasi1}
is precisely the algebra of observables of the two-body Calogero
problem.

The aim of this paper is twofold. First we stress the algebraic
aspects of our operator solution of the Calogero model, and show that
the pertinent extended Heisenberg algebra can be given a complex, (or
Bargmann-Fock type) representation in which the wave raising
operators, and thus the wave functions, become very simple. It is in
fact this representation that provides the link between the Calogero
model and anyons in the lowest Landau level.  That these two systems
are in fact equivalent, was conjectured in
\cite{hans4} and in this paper we provide the
proof.  Secondly we extend our operator method to the supersymmetric
Calogero model originally constructed by Freedman and
Mende\cite{free1}.  In fact, we show that this model is only one in a
family of (non-super symmetric) extensions of the Calogero model.
Another member of this family coincides with one of the integrable
models of Calegero type with internal degrees of freedom, recently
found by Minahan and Polychronakos\cite{minn1}.  We also construct the
relevant extended super-Heisenberg algebra, and give the Bargman-Fock
formulation of the super-Calogero model.

As already mentioned, the central ingredient of our metod is an
extended Heisenberg algebra which, in the simple case of 2 particles
(eliminating the center of mass coordinate), is defined by,
\be
&[D\,,x]=1-\nu K\, \\ &KD=-DK, \\ &Kx=-xK, \\ &K^2 =1 \ \ \ \ \ ,
\ee
where $K$ is the so called Klein operator and $\nu$ is an arbitrary
constant.

As emphasized in \cite{vasi5,vasi1} the higher-spin algebras used in
\cite{vasi2,vasi3} amount to algebras of functions of $x$, $D$, and
$K$ (\ie the algebras with generating elements $x$, $D$, and $K$
subject to (1.1) - (1.4)).  It was also shown in \cite{vasi5,vasi1}
that bilinears of $x$ and $D$, i.e. $x^2$, $D^2$, and $\{x,D\}$ span
the Lie algebra $sl_2$ with respect to commutators, while its
quadratic Casimir operator depends on $\nu$. In the group theoretical
formulation of the two-body Calogero model, with $sl_2$ as a spectrum
generating algebra\cite{pere2}, $\nu$ is identified with the Calogero
coupling constant.  The same relation was also found in the two-anyon
case and this observation was in fact the starting point for our
analysis.

Remarkably enough, (1.1) can be consistently generalized to the
$N$-body case \cite{poly6,brin3}
\be
&[D_i \,,x_j ]= A_{ij} =\delta_{ij} (1+\nu \sum_{l=1}^N K_{il})-\nu
K_{ij}, \\ &[ x_i , x_j ] = 0, \\ &[ D_i , D_j ] = 0,
\ee
where $K_{ij}$ are generating elements of the symmetric group,
\be
&K_{ij}x_j = x_i K_{ij}\ \\ &\quad K_{ij}D_j = D_i K_{ij}\,,\quad
\ee
\be
&K_{ij} =K_{ji}\ \\ &\quad K_{ij} K_{jl}=K_{jl}K_{li} =K_{li}K_{ij}
\ee
and all quantities $K_{ij}$, $x_l$, and $D_k$ are mutually commuting
when all the labels like $i,j,l,$ and $k$ are pairwise noncoinciding.
Let us emphasize that the center of mass coordinates
\be
\tilde{x}=\sum_{i=1}^N x_i\,,\qquad \tilde{D} =\sum_{i=1}^N D_i
\ee
decouple from (\ie commute with) the relative coordinates, and that
the two-body relations (1.1) - (1.4) hold for the relative coordinates
$x=x_1 -x_2$, $2D=D_1 -D_2$, $K=K_{12}$.

As will be demonstrated in section 2, the algebra (1.5) - (1.7) is
crucial for understanding the $N$-body Calogero model. Raising and
lowering operators can be constructed as $a^\pm_i = (x_i \mp
D_i)/{\sqrt{2}}$, and the generators of the spectrum generating $sl_2$
algebra (which gives the complete solution to the two-body problem)
can be constructed from bilinears in these step operators.

Several problems of both physical and mathematical nature remain to be
investigated.  On the physics side, one should ask in particular
whether there is any interesting theory like \eg some generalized
higher-spin theory or string theory with an underlying
infinite-dimensional symmetry generated by the relations (1.5) -
(1.7).

An interesting mathematical issue is the relation to the spherical
function theory on a Riemanian symmetric spaces, namely to the
analysis of hypergeometric functions associated with a root systems,
and particularly the connection to the so called "Dinkham shift
operators" which were introduced in
\cite{heck1,opda1}.\footnote{
We are grateful to Dr. A. Matsuo for informing us about these works.}
They turn out to be closely related to the realization of the
operators $D_i$ given in the section 2.  These operators are a
particular case of differential-difference operators introduced in
\cite{dunk1} for arbitrary Coxeter group, which corresponds to the
root system $A_{N-1}$.

\vskip 1cm\noi
\renewcommand{\theequation}{2.\arabic{equation}}
\setcounter{equation}{0}
\noi {\large\bf 2. Operator structure of the Calogero model}
\vskip 3mm \noi

In this section we first recapitulate the operator solution to the
Calogero problem given in \cite{brin3}. The model is defined by the
Hamiltonian
\be
H_{Cal} =\half \sum_{i=1}^N \left[ -d_i^2 + x_i^2 \right] +
\sum_{j < i}^N \frac g {(x_i-x_j)^2} \ \ \ \ \ ,
\ee
(where $d_i=\frac \partial {\partial x_i}$) which differs from that in
reference \cite{calo2} by an overall normalization and a harmonic
oscillator term for the center of mass coordinate, and we have also
put the frequency $\omega$ to one.  This must be borne in mind when
explicitly comparing our spectrum with that of \cite{calo2}.

To solve $H_{Cal} \Psi = E\Psi$, we make the ansatz:
\be
\Psi^\pm = \prod_{i>j}
(x_i -x_j)^{\nu} \Phi^\pm = \beta^{\nu}\Phi^\pm \ \ \ \ \ ,
\ee
where $x_i >x_j $ for $i\,>\,j$, while $+$ and $-$ refers to totally
symmetric and antisymmetric wave functions $\Phi^+$ and $\Phi^-$,
respectively.  Consider now the deformed Heisenberg algebra (1.5) -
(1.7). A representation of the "covariant derivative" $D_i$ can be
given as \cite{brin3}
\be
 D_i =d_i + \nu \sum_{j\neq i} \frac 1 {(x_i -x_j) } (1-K_{ij} ).
\ee
(Note that essentially the same modified derivatives were introduced
previously in \cite{dunk1}. There the operators $K_{ij}$ were not used
explicitly but rather their representation on the functions of $x_i$.
A slightly different expression for $D_i$ with the factor of $K_{ij}$
instead of $(1-K_{ij})$ was given in \cite{poly6}. The definition
(2.3) is unambiguously fixed by the requirement that $D_i$ induces no
new poles when acting on regular functions of $x_i$. In particular
$D_i$ can easily be seen to leave the space of polynomials invariant.)

The square of this operator is then \be D^2 = \sum_{i=1}^N D_i^2 =
\sum_{i=1}^N \left[ d_i^2 + \nu \sum_{i\neq j} \frac 2 {x_i - x_j} d_i - \nu
     \sum_{i\neq j} \frac 1 {(x_i-x_j)^2} (1 - K_{ij}) \right] \ .
\ee
Noting that $(1-K_{ij})$ gives 0 and 2 respectively on the totally
symmetric and antisymmetric wave functions, we finally get \be
H_{Cal}\Psi^\pm = \beta^{\nu} \half (-D^2 + X^2) \Phi^\pm \ee where
$X^2 = \sum_{i=1}^N x_i^2$, and $g=\nu(\nu\mp 1 )$, where the upper
and lower sign refers to symmetric and antisymmetric wave functions
respectively.

Given the complicated form of the $D_i$:s the algebra (1.5) - (1.7) as
well as (2.5) is amazingly simple.  Note that the $A_{ij}$ in (1.5) is
symmetric in $i$ and $j$ so that we can construct creation and
annihilation operators via \be a_i^\mp =\frac 1 {\sqrt 2} (x _i \pm
D_i ) \ee obeying the commutation relations \be [a_i^\pm ,a_j^\pm] &=&
0 \nonumber \\ \left[a_i^- ,a_j^+\right] &=& A_{ij} \ \
\ \ \ .  \ee
The Hamiltonian can now be expressed as \be H= \half (-D^2 + X^2)
=\frac{1}{2}\sum_i \{a_i^+ ,a_i^- \} \ \ \ \ \ , \ee and turns out to
obey the standard commutation relations with the creation and
annihilation operators, \be [H,a_i^\pm ] = \pm a_i^\pm \ \ \ \ \ .
\ee This last relation again follows from a series of nontrivial
algebraic manipulations using the properties of the $K_{ij}$.  The
eigenfunctions are now obtained via the construction,
\be
\Phi^\pm(n_i) =  {\cal {S}}\{ \prod_{i=1}^N (a_i^+)^{n_i}\} \Phi_0^\pm
                        \ \ \ \ \ ,
\ee
where ${\cal S }$ denotes total symmetrization, and the vacuum state
$\Phi_0^\pm$ satisfies \be a^-_i \Phi_0^\pm = 0 \ \ \ \ \ ,
\ee
and
\be
 K_{ij}\Phi_0^\pm =\pm \Phi_0^\pm \ \ \ \ \ .
\ee
It is well known that every symmetric polynomial of $a^+_i$ or $a^-_i$
can be expressed as a polynomial of elementary symmetric polynomials
\be
C_n^\pm = \sum_{i=1}^N \,(a^\pm_i)^n\,,
\ee
hence (anti)symmetric wave functions have the following basis
functions
\be
\Phi^\pm_{\{n_i\}} =  \prod_{i=1}^N (C_i^+)^{n_i} \Phi_0^\pm \ \ \ \ \ .
\ee

Using (2.8), (2.11), (2.12) and the commutation relations (2.7) we
find the ground state energy of $H_{Cal}$ to be $E_0^\pm = \frac N 2
\pm \nu\half N(N-1) $, so the complete spectrum is that of N bosons or
fermions in a harmonic oscillator well, shifted by this constant.
This is Calogero's original result. Solving (2.11) and (2.12) also
immediately gives Calogero's ground state wave function.  As
advertised, the new result is the explicit expression (2.10) for the
wave functions. Needless to say, the expressions very quickly become
very cumbersome because of the sums in the definition of $D_i$.

Let us now discuss the relation between our construction and the
algebraic approach of Perelomov\cite{pere3,olsh2} and
Gambardella\cite{gamb1}.  To this end it is convenient to use the
center mass coordinate frame.  After separating the center of mass
coordinates our construction can be rewritten in terms of the relative
derivatives $\tilde{D}_i$:
\be
\tilde{D}_i = D_i - \frac 1 N \sum_{l=1}^N d_l \ \ \ \ \ .
\ee
Note that $\sum d_i = \sum D_i$ and hence $\sum \tilde{D}_i =0$. The
relative parts of creation and annihilation operators can be defined
in a similar way:
\be
\tilde{a}^\pm_i =
a^\pm_i -\frac 1 N \sum_{l=1}^N a^\pm_l \nonumber \ \ \ \ \ .
\ee
The characteristic property of these relative operators is that they
leave the space of smooth functions vanishing on the hyperplane
$\sum_1^N x_i =0$ invariant.

In \cite{pere3} Perelomov proposed to look for such symmetric (\ie
transforming (anti)symmetric wavefunctions into (anti)symmetric ones)
operators $B_n^\pm$ that increase (decrease) the energy \ie
\be
[H, B_n^\pm] = \pm n B_n^\pm
\ee
and
\be
B_n^- \Phi_0 = 0 \ \ \ \ \ .
\ee

If such operators are mutually commuting and linearly independent the
wavefunctions
\be
\Phi_{\{n_i\}} = \prod_i (B_i^+)^{n_i} \Phi_0
\ee
with $\sum_i in_i = m$ form linearly independent states degenerate in
energy.  The dimension of this subspace for a given $m$ coincides with
the number of solutions of the equation $\sum_i in_i = m$ in
non-negative integers.

In \cite{pere3} Perelomov has obtained the operators $B_2^+$, $B_3^+$
and $B_4^+$, and in \cite{gamb1} Gambardella constructed $B_5^+$. All
these operators were shown to be mutually commuting.  To construct
such operators Perelomov deduced the equations for the coefficients of
their expansion in the powers of annihilation and creation operators
of pure oscillator system - $a_i\,,\, a^\dagger_i$. These equations
follow from (2.16) and depend on the potential
$V=g\sum_{i<j}(x_i-x_j)^{-2}$. The operators $B_n^\pm$, which because
of (2.16), generate classes of states for the general N body problem,
provide the full solution for $N \leq 5$.

{}From the commutation relation (2.9) it immediately follows that
\be
[H, (\tilde{a}^\pm_i)^n] = \pm n(\tilde{a}^\pm_i)^n
\ee
Thus the elementary symmetric polynomials $C_n^\pm$ which generate all
symmetric polynomials of $\tilde{a}_i^\pm\,$ satisfy $(2.19)$ like
$B_n$ and leave the subspace of symmetric wave functions invariant.
All polynomials $C_n^+$ commute among themselves and with the
operators $K_{ij}$ .  These polynomials can be expressed in the
following form,
\be
C_n^+ = \sum_\alpha{\cal C}_n^\alpha {\cal F}_n^\alpha(K_{ij}) \ \ \ \
\ ,
\ee
where the operators $ {\cal C}_n^\alpha $ depend on $x_i$ and $\frac d
{dx_i}$ but not on the permutation operators $K_{ij}$, while the
operators ${\cal F}_n^\alpha$ are some functions of $K_{ij}$
independent of $x_i$ and $\frac d {dx_i}$.  The point is that the
operators
\be
^S C_n^+ = \sum_\alpha{\cal C}_n^\alpha {\cal F}_n^\alpha\left .\right
|_{K_{ij} =1}
\ee
are precisely Perelomov's operators $B_n^+$ because they obey all
requirements imposed in \cite{pere3}.

Actually, it can be shown that the operators $^S C_n^+$ are symmetric
(\ie commute with $K_{ij}$) and satisfy the relations $ [H\,,\,^S
C_n^+]= n\,^S C_n^+$ .  Hence the coefficients of expansion of these
operators $^SC^\pm_i$ in pure oscillator creation and annihilation
operators satisfy the equations obtained by Perelomov where in place
of potential $V=g\sum_{i<j}(x_i-x_j)^{-2}\,$ the "potential" $\tilde
{V}=-\nu/2\sum_{i\neq j}(x_i-x_j)^{-1}\,(d_i-d_j)\,$ is used. This
implies that the operators $^S C^\pm_i$ are equivalent to the
operators $B^\pm_i$ after performing the similarity transformation
(2.2).

To prove that the $^S C_n^+$:s are mutually commuting it is convenient
to use another representation of these operators,
\be
C_n^+ =^SC_n^+ \,+\,\sum_{i,j}\Phi_{ij}\cdot (1-K_{ij}) \ \ \ \ \ ,
\ee
where $\Phi_{ij}$ are some functions of $x\,,\frac d {dx}$ and
$K_{ij}$.  Because the $ C_n^+ $:s are commuting and the $^SC_n^+ $:s
are symmetrical it follows from (2.22) that
\be
0=\left[C_n^+\,,C_m^+\right]=\left[^SC_n^+\,,^SC_m^+\right]\,+
\,\sum_{i,j}\Psi_{ij}(1-K_{ij})
\ee
with some operators $\Psi_{ij}$. This implies that the symmetric
operators $^SC_n^+$ are commuting when restricted to the space of
symmetric functions.  A slightly more complicated analysis shows that
all these commutators vanish identically.

Let us mention that the above construction makes it less mysterious
why the construction of Perelomov leads to normalizable wave functions
despite the poles in the operators $B^+_n$.

It is also to be noted, that the operators $B_2^\pm$ together with
$B_2^0 = \half H$, satisfy the $sp(2,R)$ algebra,\footnote{Let us note
that the same algebra is often denoted as $sp(1,R).$ In particular,
such a convention was used in \cite{hans4}.}
\be
&\left[B_0, B_\pm \right] \,\,=\,\, \pm B_\pm
\ee
\be
&\left[B_+,B_-\right] \,\,=\,\, -2B_0 \ \ \ \ \ , \label{algebra}
\ee
which is the spectrum generating algebra for the $N=2$ case.  In the
following we shall consider other representations of this algebra.

\vskip 1cm
\renewcommand{\theequation}{3.\arabic{equation}}
\setcounter{equation}{0}
\noi {\large\bf 3. The Complex Representation and Anyons}
\vskip 3mm

The connection between the Calogero problem and fractional statistics
in 1+1 dimension was first noted in \cite{lein3}, and the N-body
problem was discussed in \cite{poly4} and \cite{hans4}.  In particular
it was noted that there is a close similarity between the Calogero
problem and N anyons in the lowest Landau level\cite{hans4,poly5}.  In
\cite{hans4} it was shown that the system of two anyons in the lowest
Landau level is in fact equivalent to the 2-body Calogero problem. It
was also shown that, after appropriate rescalings, the spectrum of the
total angular momentum operator for N anyons is identical to that of
the N-body Calogero Hamiltonian.  The wave functions for anyons in the
lowest Landau level are all explicitly known, and can in fact be
constructed with the help of raising and lowering operators. The
conjecture in \cite{hans4} that the systems are in fact equivalent,
was strongly supported by our operator construction of the wave
functions. We shall now prove this by finding a complex representation
of the operator algebra (1.5) - (1.7) and showing that the
corresponding states are precisely those of anyons in the lowest
Landau level. In the case of two particles, we also explicitly show
the connection to the treatment in \cite{hans4}.

\newcommand\zbar {\overline{z} }

We start from the Bargmann-Fock representation for a collection of
harmonic oscillators, where the creation and annihilation operators
are represented by
\be
\tilde{a}_i &=& \frac \partial {\partial z_i} = \partial_i  \\
\tilde{a}_i^{\dagger} &=& z_i \ \ \ \ \ ,
\ee
where $z_i= \frac 1 {\sqrt{2}} (x_i + i p_i)$. These operators act on
holomorphic functions of $z$, and the scalar product (in the one
particle case) is defined by,
\be
\langle \psi_2 |\psi_1 \rangle = \int \frac {dz d\overline{z} } {2\pi i}
e^{-z\zbar} \, \overline{\psi_2 (z)} \psi_1(z) \ \ \ \ \ ,
\ee
where bar denotes complex conjugation.  In this representation the
$g=0$ part of the Calogero Hamiltonian (2.1), \ie a system of
particles interacting only by harmonic forces, can be written
\be
H_{ho} = \sum_{i=1}^N (z_i\partial_i + \half) \ \ \ \ \ .
\ee
An ON basis is given by products of the one-particle states
\be
\psi_{n_i}(z_i) = \frac 1 {\sqrt{n_i!}}\, z_i^{n_i}  \ \ \ \ \ .
\ee
This representation for the harmonic oscillator is connected to the
conventional one (where the wave functions involve Hermite
polynomials) via a unitary transformation. On the other hand, we can
interpret $H_{ho}$ as the angular momentum operator of N particles in
the lowest Landau level in radial gauge ($A_r=0$). More precisely
$H_{ho} = \sum_{i=1}^N (\hat L_i + \half)$, where $\hat
L_i=z_i\partial_i$ is the angular momentum operator of particle $i$.

We now introduce the operators
\be
{\tilde a}^-_i &=& \partial_i + \nu \sum_{j\neq i} \frac 1 {(z_i
-z_j)} (1-K_{ij} )\\ {\tilde a}^+_i &=& z_i \ \ \ \ \ .
\ee
These are direct complex generalizations of the operators $D_i$ and
$x_i$ in section 1 and satisfy the same algebra (1.5) - (1.7). The
Hamiltonian (2.8) generalizes to
\be
\hat H = \sum_{i=1}^N \half \{ {\tilde a}^-_i  , {\tilde a}^+_i \} =
     \sum_{i=1}^N \left[ z_i\partial_i + \half \right]+ \frac \nu 2
N(N-1) \ \ \ \ \ ,
\ee
which again describes particles in the lowest Landau level but with
the angular momentum shifted by a constant value $\nu$ for each of the
$N(N-1)/2$ pairs.  This is precisely a system of N anyons! Note that
although the annihilation operators, ${\tilde a}^-_i $ are
complicated, the creation operators ${\tilde a}^+_i = z$ act trivially
on the wave functions. In the real (or Calogero) representation both
$a_i$ {\em and} $a_i^\dagger$ act non-trivially on the wave functions.
This is the basic reason for why the wave functions are so simple in
the complex (or anyonic) description, and so complicated in the real
one.

For the following discussion of the $N=2$ case, we need the explicit
expressions, in the complex representation, for the operators
$B_2^\pm$ and $B_2^0$ introduced at the end of section 2. A simple
calculation yields
\be
\tilde B_2^+ &=& \half  z^2  \nonumber \\
\tilde B_2^- &=& \half D^2 = \half  \left[ \partial^2 + \frac {2\nu} z \partial
         -\frac \nu {z^2} (1-K) \right] \label{comprep} \\
\tilde B_2^0 &=& \half \left[\hat L + \half \right] =
      \half \left[ z\partial + \nu + \half \right] \ \ \ \ \ ,
\nonumber
\ee
where $z=z_1 - z_2$ is the relative coordinate, $\partial$ and $D$ the
corresponding derivatives, and $K=K_{12}$. The operators are properly
normalized to satisfy the sp(2,R) algebra (2.24), (2,25).

We are now ready to show how the above results relate to those in ref.
\cite{hans4}. There
the equivalence between the two-body Calogero system and two anyons in
the lowest Landau level was demonstrated by an explicit construction
of the generators for the spectrum generating sp(2,R) algebra (2.24),
(2,25) in the anyon case.  For this, consider the representations in
the discrete series defined by,
\be
B_0|k,\mu\rangle &=& (k+\mu)\, |k, \mu \rangle \label{Ak} \\
\Gamma|k,\mu\rangle &=& \mu(\mu-1)\, |k,\mu \rangle \ \ \ \ \ ,
\label{gamma}
\ee
where $\Gamma = B_0^2 -\half(B_+B_- + B_-B_+)$ is the quadratic
Casimir operator, $\mu > 0$ and $k=0,1,2..$. Different values for the
real parameter $\mu$ correspond to inequivalent representations. With
appropriate phase conventions the commutation relations (2.24), (2,25)
imply
\be
B_+|k,\mu\rangle &=& \sqrt{(k+1)(k+2\mu)}\,|k+1,\mu\rangle
\label{Bplus} \\
B_-|k,\mu\rangle &=& \sqrt{k(k+2\mu-1)} \,|k-1,\mu\rangle \ \ \ \ \ .
\label{Bminus}
\ee
Using these relations, it was shown in \cite{hans4} that the $B_\pm$
can be represented by the following differential operators,
\be
\hat B_2^0 &=& \half\left(z\hat D+\half\right)  \nonumber \\
\hat B_2^+ &=& \half z^2 M  \label{oldrep}   \\
\hat B_2^- &= & \half M \hat D^2  \nonumber \ \ \ \ \ ,
\ee
where $\hat D = \partial + \bar z /2\,\,$ \footnote{ Note that in
\cite{hans4} the exponential factor was included in the wave functions
rather than in the measure, hence the occurrence of $\hat D$ rather
than $\partial$.} and $M$ is the square root of the operator
\be
M^2 = 1-\frac {\nu(\nu-1)} {( L +1)( L +2)} \equiv 1-\nu(\nu-1) \frac
1 {z^2} \frac 1 {\hat D^2} \ \ \ \ \ ,
\ee
which is positive and hermitian, and where $\mu$ in (\ref{Bminus}) and
(\ref{Bplus}) is related to the anyonic parameter $\nu$ by
\be
\mu = \frac {\nu}{ 2} + \frac 1 4  \label{mu} \ \ \ \ \ .
\ee
The operators (3.14) act on the normalized ''anyonic'' wave functions
in the lowest Landau level given by,
\be
\psi^\nu_k =  N_{2k+\nu} z^{2k+\nu} e^{-\half \overline z z }
      \ \ \ \ \ , \label{LLLan}
\ee
where $N_{\ell } = [\pi\Gamma(\ell+1)]^{-\half} \label{LLL} $, and the
integer $k$ is restricted by the requirement that the angular momentum
${\ell} = 2k+\nu $ fulfills $\ell \ge 0$ in order for the wave
function to be regular at the origin.

Thus, the observables (\ref{oldrep}) for the two anyon problem,
satisfy the same algebra as the observables in the $N=2$ Calogero
problem, and in
\cite{hans4} it was concluded that the systems are indeed equivalent.
This is of course nothing but a special case of the equivalence just
proved for general N.  It remains to understand the connection between
the representation (3.14) found in \cite{hans4} and the representation
(3.9) obtained from our creation and annihilation operator formalism.
We do this by following the same procedure as in \cite {hans4}, but
{\em require} that $\tilde B^+ = \half z^2$ and $\tilde B^0 = \half
(z\partial + \half +\nu)$ and act on the functions
\be
\tilde \psi_k (z)  = \tilde N_k z^{2k} \ \ \ \ \  .
\ee
{}From (\ref{Bplus}) it now follows
\be
\tilde N_{k-1} = 2\sqrt{k(k-\half +\nu)}\,  \tilde N_k  \ \ \ \ \  .
\ee
We then make the ansatz,
\be
\tilde B^- = \half \tilde M \partial^2 \ \ \ \ \  ,
\ee
and use (3.19) and (3.13) to obtain
\be
\tilde M = 1 + \frac {2\nu} {2k-1} = 1 + \frac 1 z \frac 1 \partial
      \ \ \ \ \ ,
\ee
so finally
\be
\tilde B^- = \half \partial^2 + \nu \frac 1 z \partial \ \ \ \ \   ,
\ee
which is exactly the result (\ref{comprep}) obtained by squaring our
creation operator for the symmetric case where $K=1$.  We thus see
that there is a large freedom in representing the algebra (2.24),
(2,25) on holomorphic functions, and we have explicitly demonstrated
how to construct the representation (\ref{comprep}) both in our
approach, and in that used in \cite{hans4}.

Note that $\tilde B_2^-$ in (\ref{comprep}) annihilates both the even
and the odd ground states (\ie both 1 and z), while $\tilde B^-$ in
(3.22) only annihilates the even one.  This is because in the latter
construction we explicitly used the symmetric functions (3.18). It is
easy to repeat the construction for wave functions $\sim z^{2k+1}$ to
obtain the extra term $\sim 1/z^2$ in (\ref{comprep}) for the odd
case.

\renewcommand{\theequation}{4.\arabic{equation}}
\setcounter{equation}{0}
\vskip 1cm \noi
\noi {\large\bf 4. Spinning Models and the Super-Calogero Model}
\vskip 3mm
In section 2 we saw that the spectrum of the Calogero model coincides
with the one of the ordinary harmonic oscillators, apart from a shift
of the vacuum energy. It is also known that the one-dimensional
harmonic oscillator has a unique, $N=2$, supersymmetric extension.
This is easily seen in a Lagrangian formulation of the one-particle
case. Define the superfield
\be
\Phi(\tau,\vartheta) = x(\tau) + i\vartheta^i \theta^i(\tau) + i
\vartheta^1\vartheta^2 F(\tau) \ \ \ \ \ ,
\ee
and construct the covariant derivative
\be
{\cal D}_i = \frac {\partial}{\partial \vartheta^i} + i\vartheta^i
\frac {\partial}{\partial\tau} \;\;\; i = 1,2.
\ee
Then consider the action
\be
S = -\frac {1}{2}\int \!d\tau d\vartheta^1 d\vartheta^2 [{\cal D}_1
\Phi{\cal D}_2 \Phi + i \omega \Phi^2] \ \ \ \ \ .
\ee
On dimensional grounds we see that this construction is unique.  Note
that we have reintroduced the frequency $\omega$ to make the argument
about the dimension clearer. It will also be useful to have it
explicit for the discussions that will follow. If we perform the
$\vartheta$-integrations, eliminate the auxiliary fields and go over
to a hamiltonian formalism with N interacting particles and set
\be
&a_i= \frac{1}{\sqrt2}(\omega x_i +i p_i) \\ &a_i^\dagger =
\frac{1}{\sqrt2}(\omega x_i-i p_i) \\ &\theta_i = \frac{1}{\sqrt2}
(\theta_i^1 + i \theta_i^2) \\ &\theta_i^\dagger = \frac{1}{\sqrt2}
(\theta_i^1 -i \theta_i^2)
\ee
and construct the supercharges
\be
&Q = \sum_i \theta_i^\dagger a_i \\ &Q^\dagger = \sum_i
\theta_ia_i^\dagger,
\ee
then the quantized Hamiltonian takes the following explicitly
supersymmetric form
\be
H =\{ Q, Q^\dagger \} = \half \sum_i \left( \{a_i^\dagger, a_i\} +
\omega [\theta_i^\dagger, \theta_i]\right) \ \ \ \ \  .
\ee

The wave function for the $N$ particle system is a sum of the form
\be
\Psi(x_m,\theta_i) = \psi(x_m) + \sum_i \psi_i (x_m)\theta_i +
   \sum_{ij}\psi_{ij}(x_m)\theta_i\theta_j + .......\nonumber
\ee
To describe a system of identical bosons and fermions, we demand that
$\Psi$ is symmetric or antisymmetric under the combined exchange
$(x_i, \theta_i) \leftrightarrow (x_j,
\theta_j)$. It is easy to see that this will ensure the correct symmetrization
and antisymmetrization of the bosonic and fermionic wavefunctions
respectively.

It is now straightforward to supersymmetrize the Calogero model by
simply replacing the free raising and lowering operators $a_i^\dagger$
and $a_i$ with the corresponding operators $a_i^+$ and $a_i^-$ given
in (2.6). Note that since the frequency is reintroduced, the
commutation relations (2.7) take the form $[a_i^-, a_j^+]=\omega
A_{ij}$. The Hamiltonian corresponding to $H$ in (2.8) is
\be
H_s = \{ Q^+, Q^- \} = \half \sum_i \{a_i^+, a_i^-\} + \half \omega
\sum_{i,\, j} [\theta_i^\dagger, \theta_j] A_{ij} \ \ \ \ \ .
\ee
Here
\be
&Q^- = \sum_i \theta_i^\dagger a_i^- \\ &Q^+ = \sum_i \theta_ia_i^+
\ee
By introducing the explicit form (1.5) for the $A_{ij}$:s, the last
term in (4.11) becomes
\be
\half  \omega \sum_{i,\, j} [\theta_i^\dagger, \theta_j] A_{ij}  =
\half  \omega\sum_i [\theta_i^\dagger, \theta_i] +
\frac {\nu\omega} 2
\sum_{j\neq i} \half  [\theta_i^\dagger -\theta_j^\dagger, \theta_i-\theta_j]
K_{ij}
\ee
By direct calculation, one can now show that the combination
\be
K^\theta_{ij} = \half [\theta^\dagger_i -\theta^\dagger_j ,\theta_i -
\theta_j]= 1-(\theta_i - \theta_j)(\theta^\dagger_i -\theta^\dagger_j)
\ee
occurring in (4.14) is the permutation $K$ operator for the fermionic
variables,
\ie it fulfills
\be
\theta_i K^\theta_{ij} =
     K^\theta_{ij} \theta_j \,\,\,\,,\,\,\,\,\,\,(K^\theta_{ij})^2=1
\ee
and the algebraic relations (1.10) and (1.11). A slightly different
realization of the permutation operators acting on inner labels was
used in the paper by Minahan and Polychronakos\cite{minn1}. We see
that $K^\theta_{ij}$ naturally appear in the supersymmetric extension
of the Calogero model.  Since $K_{ij}$ and $K^\theta_{ij}$ commute we
can define a total $K$ operator, again satisfying (1.8) - (1.11), that
exchanges both bosonic and fermionic coordinates by
\be
K^{tot}_{ij} = K^{}_{ij} K^{\theta}_{ij} \ \ \ \ \ .
\ee
and rewrite the Hamiltonian (4.11) as
\be
H_s = \half \sum_i \{a_i^+, a_i^-\} + \half \omega \sum_{i}
[\theta_i^\dagger, \theta_i] + \half \omega \nu \sum_{i\neq
j}K_{ij}^{tot} =H_B+H_F+H_K\, .
\ee

Now one observes that when restricted to the subspaces of totally
symmetric or totally antisymmetric wavefunctions the operators
$K^{tot} _{ij}$ act as a constant, $K^{tot} _{ij} = \pm 1, $ and,
therefore, for these subspaces the Hamiltonian (4.18) amounts to a
particular case of the family of Hamiltonians
\be
H = \half \sum_i \{a_i^+, a_i^-\} + \half \omega_F \sum_{i}
[\theta_i^\dagger, \theta_i] + c
\ee
with $\omega_F=\omega$ and $c=\pm \frac {\nu\omega} 2 N(N-1)$.

Evidently the Hamiltonian (4.19) obeys the following commutation
relations $$
\left[H,a^\pm_i\right]=\pm\omega a^\pm_i \,,\qquad
\left[H,\theta^\dagger_i\right]=\omega_F \theta^\dagger_i \,,\qquad
\left[H,\theta_i\right]=-\omega_F \theta_i
$$ and therefore is exactly solvable for arbitrary values of bosonic
and fermionic frequencies.  All eigenstates of (4.19) can be obtained
by acting with creation operators $a^+_i$ and $\theta^\dagger_i$ on
the groundstate $\Phi_0$ which satisfies the equations $$
a_i\Phi_0=\theta_i\Phi_0=0 $$

Groundstates have the following form $$
\Phi_0(x,\theta)= \theta_1 \theta_2 \, ... \,\theta_N \Phi_0(x).
$$ Here $\Phi_0(x,\theta)$ as well as $\Phi_0(x)$ has a definite
parity
\be
K_{ij}^{tot}\Phi_0(x,\theta)=\pi_0\Phi_0(x,\theta),\qquad
K_{ij}^{tot}\Phi_0(x)=-\pi_0\Phi_0(x),
\ee
where $\pi_0=\pm 1$.  The ground energy can be easily computed
\be
E_0=\half N(\omega-\omega_F)+c-\pi_0 \frac {\nu\omega} 2 N(N-1) \ \ \
\ \ ,
\ee
and as expected $E_0=0$ in the supersymmetric case.  The symmetric and
antisymmetric excited eigenstates of (4.19) can be obtained by acting
on a symmetric or antisymmetric $\Phi_0$ with polynomials of $a^+_i$
and $\theta_i^\dagger$ that are symmetric under the total exchange
$K_{ij}^{tot}$.

Using (4.15) - (4.17) one can write $K_{ij}$ as $K_{ij}^\theta
K_{ij}^{tot}$. When restricted to the subspace of symmetric
wavefunctions with $K_{ij}^{tot}=1$ this enables one to replace
bosonic permutation operators with the fermionic ones that effectively
leads to interactions between bosons and fermions.  For the particular
case of the supersymmetric Hamiltonian (4.11) we reintroduce the
factor $\beta^\nu$ to get the following supersymmetric generalization
of the Hamiltonian in (2.1)
\be
H_{sCal} = \half \sum_i \left[-d_i^2 + \omega^2 x_i^2\right] +
\sum_{i<j}\frac {\nu^2} {(x_i-x_j)^2} + \half \omega\sum_i
[\theta_i^\dagger, \theta_i] \nonumber \\ +\frac \nu 4 \sum_{i\neq j}
\frac 1 {(x_i - x_j)^2}[\theta_i - \theta_j , \theta^\dagger_i
-\theta^\dagger_j ] +\frac {\nu\omega} 2 N(N-1) \ \ \ \ \ .
\ee
This is exactly the supersymmetric extension of the Calogero model
originally found by Freedman and Mende\cite{free1}.

Some comments are now in order. It is clear that supersymmetry is not
important for solvability of these models. Moreover, because of the
term $H_K$ in (4.18) the supersymmetric model turns out to be
explicitly solvable only when restricted to the subspace of totally
(anti)symmetric wavefunctions when $H_K$ amounts to some constant. On
the other hand the Hamiltonians (4.19) are explicitly solvable for
arbitrary $\omega_B$ and $\omega_F$ and all types of symmetry
properties of wave functions.

It is also clear that if we introduce an internal symmetry by adding
more fermionic variables $\theta^a_i,\ \ a=1,2,..$ we can generalize
(4.19) to other solvable models.  For the particular case $\omega_F=0$
these Hamiltonians coincide with a special case of those considered by
Minahan and Polychronakos in \cite{minn1}.

It is also amusing to write the supersymmetric version of the
''anyonic'' representation of the Calogero model, \ie the
supersymmetric version of (3.8)
\be
\hat H_{sCal} = \half  \sum_i\left[ \{z_i , \frac \partial {\partial z_i} \}
      - [\theta_i , \frac \partial {\partial \theta_i} ] \right] \ \ \
\ \ .
\nonumber
\ee
Although written in this way the supersymmetric Calogero model indeed
looks rather trivial, the unitary transformation that relates the two
representations to each other is not trivial at all.

As shown by Freedman and Mende\cite{free1}, the super Calogero model
possesses a dynamical $OSp(2;2)$ supersymmetry. We will now show how
this symmetry can be realized in terms of bilinears of the modified
creation and annihilation operators that we already introduced.  The
result is quite simple since all basic quantities are expressed in
terms of the modified creation and annihilation operators essentially
in the same way as for the ordinary harmonic oscillator.  (To simplify
formulae from now on we again set $\omega =1.$)

The basic supercharges have already been introduced in (4.12) and
(4.13), and they of course fulfill
\be
(Q^+)^2 =(Q^-)^2 =0\,,\nonumber
\ee
We also introduce the notation
\be
 \{Q^+,Q^-\} = {H} = T_3 +J
\ee
where
\be
T_3 = \frac{1}{2} \sum_i \{a^-_i \,,a^+_i\}
\ee
\be
J=\frac{1}{2} (\sum_i [\theta^+_i \,,\theta_i^- ] +\nu \sum_{i\neq j}
K^{tot}_{ij}) \ \ \ \ \ ,
\ee
where $\theta^+ =\theta^\dagger$ and $\theta^- =\theta$.  To obtain
the full $osp(2;2)$ algebra we need the additional supercharges \be
S^- =\sum_i a^-_i \theta^-_i\,,\qquad S^+ =\sum_i a^+_i
\theta^+_i\,,\qquad \ee and the bosonic charges
\be T_\pm = \half  \sum_i (a^\pm_i)^2 \ee
The bosonic operators $T_\pm $, $T_3 $ and $ J$ span the algebra
$sp(2) \oplus o(2)$, the bosonic subalgebra of $OSp(2;2)$, \be [T_3
,T_\pm ] =\pm2T_\pm\,,\qquad [T_- ,T_+ ] = T_3
\ \ \ \ \ ,
\ee and $J$ commutes with all bosonic generators.

The nontrivial anticommutators are
\be
\{S^+,Q^+\} = T_+ \,,\qquad \{ Q^-,S^-\} = T_- \\
\{Q^+ ,Q^-\} =  T_3 +J\,,\qquad
\{S^+ ,S^-\} =  T_3 -J\,,\qquad
\ee
(all other anticommutators vanish). The nonvanishing boson-fermion
commutators read
\be
  &[T_+ ,Q^- ] =-S^+ \,,\qquad [T_+ ,S^-] =- Q^+\\ &[T_- ,S^+ ] = Q^-
\,,\qquad [T_- ,Q^+] = S^- \\ &[T_3 ,Q^+] = Q^+ \,,\qquad [T_3 ,Q^-] =
-Q^- \\ &[T_3 ,S^-] =-S^- \,,\qquad [T_3 ,S^+] = S^+ \\ &[J ,S^-]
=-S^- \,,\qquad [J , S^+] = S^+ \\ &[J ,Q^+] =-Q^+ \,,\qquad [J , Q^-]
= Q^- \ \ \ \ \ .
\ee

One can rewrite this in a more systematic way by introducing
supercharges
\be
Q^{\alpha b}=\sum_i a^{\alpha}_i \theta^b_i
\ee
and bosonic charges
\be
B^{\alpha\beta}=\half \sum_i \left\{a^{\alpha}_i\,,\,
a^\beta_i\right\} \\ J^{a b}=\sum_{i\neq j}\left[\theta^a_i\,,\,
\theta^b_j \right] A_{ij}
\ee
with $\alpha ,\,\beta,\, a,\, b \,= \,\pm 1$.

It is seen easily that
\be
J^{a b}=\delta_{a+b\,\,0}\,\,a\,J
\ee

The commutation relations (4.28) -(4.36) can be written now as
\be
\left\{Q^{\alpha a}\,,\,Q^{\beta b}\right\}=\delta_{a+b\,\,0}\, B^{\alpha\beta}
      +\delta_{\alpha+\beta\,\,0}\,\beta\, J_{ab} \ \ \ \ \\
\left[B^{\alpha \beta}\,,\,Q^{\gamma a}\right]=
      \delta_{\alpha+\gamma\,\,0}\,\gamma \,Q^{\beta a} +
\delta_{\beta+\gamma\,\,0}\,\gamma\, Q^{\alpha a} \ \\
\left[B^{\alpha \beta}\,,\,B^{\gamma \chi}\right]=
      \ \ \ \ \ \ \ \ \ \ \ \ \ \ \ \ \ \ \ \ \nonumber \\
\delta_{\alpha+\gamma\,\,0}\,\gamma \,B^{\beta \chi}+
\delta_{\alpha+\chi\,\,0}\,\chi\, B^{ \beta\gamma }+
\delta_{\beta+\gamma\,\,0}\,\gamma \,B^{\alpha \chi}+
\delta_{\beta+\chi\,\,0}\,\chi \,B^{\alpha \gamma } \\
\left[B^{\alpha \beta}\,,\,J^{a b}\right]= 0 \ \ \ \ \ \ \ \ \ \ \ \
\ \ \ \ \
       \\
\left[J\,,\,Q^{\alpha a}\right]=\,a\,Q^{\alpha a}  \ \ \ \ \ \ \ \ \ .
\ee

\renewcommand{\theequation}{5.\arabic{equation}}
\setcounter{equation}{0}
\vskip 1cm \noi
\noi {\large\bf 5. Fermionic Covariant Derivatives}
\vskip 3mm \noi

\newcommand\cD{{\cal D}}
\newcommand\cd{{\partial}}

In the previous section we supersymmetrized the Calogero model by
adding anticommuting harmonic oscillators. We could also ask whether
we can extend the extended Heisenberg algebra to an extended
super-Heisenberg algebra. To this end, we add operators $\cD_i$ and
$\theta_i$, with the non-trivial (anti-)commutation rules,
\be
&\{ \cD_i , \cD_j \} = 2 \delta_{ij} D_i \\ &\{ \cD_i , \theta_j \} =
\delta_{ij} \\ &[\cD_i , x_j ] = \theta_i A_{ij} \ \ \ \ \ .
\ee
A representation of the algebra (5.1) - (5.3) is given by the standard
construction,
\be
\cD_i=\cd_i + \theta_i D_i
\ee
so that
\be
\cD_i = \cd_i +\theta_i d_i+ \sum_{k\neq i}\frac{{\nu\theta_i
}}{{x_i - x_k}}(1-K_{ik}) \ \ \ \ \ ,
\ee
where as previously $K_{ik}$ is the bosonic permutation operator
interchanging $x$ -coordinates and
\be
\partial_i = \theta_i^{\dagger}\,\,
\qquad d_i =\frac{d}{dx_i} \ \ \ \ \ .
\ee

{}From the general viewpoint one can consider the relations (1.5) -
(1.11), as defining relations for generating elements of a (new?)
class of infinite-dimensional associative algebras, deformed
Heisenberg-Weyl algebras. Then the relations (5.1) - (5.3) extend them
to deformed super Heisenberg-Weyl algebras. An interesting question
then is if there exist physical applications of these algebras in
connection with, say, higher-spin theories in two and more dimensions.

Since the bilinears constructed from bosonic and fermionic generating
elements were shown in the previous section to form the superalgebra
$osp(2;2) $ with respect to (anti)commutators one can expect that the
deformed super Heisenberg-Weyl algebras contain the enveloping algebra
of $osp(2;2)$ at least for sufficiently high $N$ (for the condition
that $\nu$ is interpreted as an independent Abelian generator). For
the simplest case $N=2$ it was shown in \cite{berg1} that the
corresponding associative algebra is isomorphic to the enveloping
algebra of $osp(1;2)$.

Another question that may be interesting to address is whether there
exist other representations for the derivatives $\cD_i$ and $D_i$ that
obey the crucial relation (5.1) (which in its turn guarantees that the
derivatives $D_i$ are mutually commuting in accordance with (1.7))
thus leading to other deformations of the super-Heisenberg algebra.

A direct analysis shows that there is a one parameter family of such
derivatives given by,
\be
{}^{\alpha}\cD_i& =& \cd_i +\theta_i d_i+ \sum_{k\neq
i}\frac{{\nu\theta_i -\alpha\theta_k}}{{x_i - x_k}}(1-K_{ik}) \\ && +
\alpha (\nu - \alpha )
\sum_{k,l: \, k\neq i,\,
l\neq i,\, k\neq l} \theta_i\theta_k\theta_l \frac {1}{x_i - x_k}
(1-K_{ik})
\frac {1}{x_i-x_l}(1-K_{il}) \ \ . \nonumber
\ee
where $\cD_i ={}^0 \cD_i$. The basic property
\be
\{{}^{\alpha} \cD_i \, ,\,{}^{\alpha}\cD_j \} = 2\delta_{ij}\,
^{\alpha} D_i \ \ \ \ \ .
\ee
can be shown to hold and therefore the ${}^\alpha D_i$:s are still
commuting.

Their explicit form is
\be
 {}^{\alpha}D_i &=& d_i+ \sum_{l\neq i} [\frac{\nu}{x_i-x_l}
(1-K_{il})
\nonumber \\ & &
+\frac {\alpha\theta_i \theta_l}{(x_i-x_l )^2} (1-K_{il}) +
\frac{\alpha\theta_i \theta_l}{x_i -x_l} (d_i -d_l ) K_{il} ]
\nonumber \\
&& +\nu\alpha \sum_{k,l:\,\, k\neq i,\, l\neq i, k\neq l}
\frac{\theta_i \theta_k +\theta_k \theta_l +\theta_l \theta_i} {(x_k
-x_l )(x_l -x_i )} (1-K_{il} )K_{ik} \ \ \ \ \ .
\ee
The bosonic part $H_B=\half \sum_i((-\,{}^{\alpha}D_i)^2 + x_i^2) $ of
the Hamiltonian
can be shown to have the following form
\be
&& H_B =-\half \sum_i d_i^2 + \half \sum_i x_i^2 -\half \sum_{k.l:\,k
\neq l} \{ -\frac {\nu}{(x_k - x_l )^2} (1-K_{kl})\nonumber\\ +&&
\frac {\nu}{x_k -x_l } (d_k -d_l ) -2 \frac {\alpha\theta_k \theta_l
}{(x_k -x_l )^3} (1-K_{kl}) +\frac {\alpha\theta_k \theta_l }{(x_k
-x_l )^2} (d_k - d_l ) (1-K_{kl}) \}\nonumber\\ &&
-\frac{\nu\alpha}{2} \sum_{i,k,l: \,\, i\neq k\neq l\neq i} \frac {
\theta_i \theta_k +\theta_k \theta_l + \theta_l \theta_i }{(x_i -x_k )(x_k -x_l
)(x_l -x_i )} ( \frac {1}{3} - K_{ik} + \frac {2}{3} K_{ik} K_{il} ).
\ee
Obviously these new derivatives ${}^{\alpha}\cD_i$ lead to some
solvable system and the question is if this system is really new or
this is another way for describing the original system.

The result is that all these systems are indeed pairwise equivalent.
To see this one observes that it is possible to introduce new
variables ${}^\alpha x_i$, ${}^\alpha a^\pm_i $, ${}^\alpha
\theta^\pm_i $, and $ {}^\alpha K_{ij} $
\be
{}^\alpha x_i = x_i + \alpha \sum_k \theta_i \theta_k K_{ik} \,\,,
\ee

\be
{}^\alpha a^{ \mp}_i =\frac{1}{\sqrt{2}} ({}^\alpha x_i \pm {}^\alpha
D_i )
\,\,,
\ee
\be
{}^\alpha \theta_i^- = \theta_i \,\,,
\ee
\be
{}^\alpha \theta^+_i = {}^\alpha \cD_i - \theta_i {}^\alpha D_i
\ee
\be
{}^\alpha K_{ij} = {}^\alpha K^\theta_{ij} K^{tot}_{ij}
\qquad
{}^\alpha K^\theta_{ij} =\frac{1}{2} [\, {}^\alpha \theta^+_i
-{}^\alpha \theta^+_j \,, {}^\alpha \theta_i^- -{}^\alpha \theta_j^- ]
\ee
such that the fermionic variables obey the standard commutation
relations
\be
\{{}^\alpha \theta^\pm_i \,,{}^\alpha \theta^\pm_j \} = 0\,, \qquad
\{{}^\alpha \theta^+_i \,,{}^\alpha \theta^-_j \} = \delta_{ij}
\ee
and commute with the bosonic variables ${}^\alpha a^\pm_i $, and $
{}^\alpha K_{ij} $ while the latter obey the relations analogous to
(2.7)
\be
[{}^\alpha a^\pm_i ,{}^\alpha a^\pm _j ] = 0
\nonumber
\ee
\be
[{}^\alpha a^-_i ,{}^\alpha a^+ _j ] = \delta_{ij} (1+ \nu
\sum_{l=1}^N {}^\alpha K_{il}) - \nu\, {}^\alpha K_{ij}
\ee
and ${}^\alpha K_{ij}$ behave as the permutation operators for the
bosonic variables ${}^\alpha a^\pm_i $ obeying the relations analogous
(1.8) - (1.11).

To make the relationship between the algebras with different $\alpha$
explicit let us introduce the operator $\Theta$ \be \Theta=\sum_{i,j:
\,i \neq j}
\frac{\theta_i \theta_j}{x_i - x_j} (1-K_{ij}) \ \ \ \ \ .
\ee
Then, using the Baker-Hausdorff formula,
\be
e^{-\alpha B}\, A \,\,e^{\alpha B}=
\sum_{n=0}^\infty \frac{\alpha^n}{n!} A_n \ \ \ \ \ ,
\ee
valid for any operators $A$ and $B$ where $A_0=A$ and $A_i=[A_{i-1},\,
B]$, one can prove that
\be
{}^{\alpha} A = \exp\, (\,\frac{\alpha}{2} \,\Theta)\, A\,\,
\exp\,(-\frac{\alpha}{2}\, \Theta)
\ee
for any of operators ${}^\alpha x_i$, ${}^\alpha D_i$, ${}^\alpha
\cD_i$, ${}^\alpha a^\pm_i $, ${}^\alpha \theta^\pm_i $, and $
{}^\alpha K_{ij} $

Thus the derivatives $^\alpha\cD_i$ lead to a solvable system
equivalent to (4.19). One can conjecture that the relations (1.5) -
(1.11) and (5.1) - (5.5) exhaust all nontrivial deformations of the
(super-) Heisenberg algebra involving the permutation generators.

\vfill

\end{document}